\documentclass{IEEEcsmag}

\usepackage{upmath}
\usepackage[style=ieee]{biblatex}

\usepackage{amsmath,amssymb,amsfonts}

\usepackage{multirow}
\usepackage{algorithmic}
\usepackage{graphicx}
\usepackage{subfigure} 
\usepackage{todonotes}
\usepackage{color, colortbl}
\definecolor{LightCyan}{rgb}{0.68, 0.84, 0.86}
\definecolor{Blue}{rgb}{0.81, 0.84, 0.85}
\definecolor{LightPink}{rgb}{1, 0.60, 0.55}

\usepackage[colorlinks,urlcolor=blue,linkcolor=blue,citecolor=blue]{hyperref}
\usepackage{listings}
\jvol{NA}
\jnum{NA}
\paper{NA}
\jmonth{February}
\jname{IT Professional}
\pubyear{2021}

\begin{document}


\title{Talaria: A Framework for Simulation of Permissioned Blockchains for Logistics and Beyond}


\author{Jiali Xing}
\affil{University of Pennsylvania}

\author{David Fischer}
\affil{Duke University}

\author{Nitya Labh}
\affil{The College of William and Mary}

\author{Ryan Piersma}
\affil{Duke University}

\author{Benjamin C. Lee}
\affil{University of Pennsylvania}

\author{Yu Amy Xia}
\affil{The College of William and Mary}

\author{Tuhin Sahai}
\affil{Raytheon Technologies Research Center}

\author{Vahid Tarokh}
\affil{Duke University}

\begin{abstract}
In this paper, we present Talaria, a novel permissioned blockchain simulator that supports numerous protocols and use cases, most notably in supply chain management. Talaria extends the capability of BlockSim, an existing blockchain simulator, to include permissioned blockchains and serves as a foundation for further private blockchain assessment. Talaria is designed with both practical Byzantine Fault Tolerance (pBFT) and simplified version of Proof-of-Authority consensus protocols, but can be revised to include other permissioned protocols within its modular framework. Moreover, Talaria is able to simulate different types of malicious authorities and a variable daily transaction load at each node. In using Talaria, business practitioners and policy planners have an opportunity to measure, evaluate, and adapt a range of blockchain solutions for commercial operations.
\end{abstract}

\raggedbottom


\maketitle

\clearpage

\section{Introduction}

\subsection{The Emergence of Blockchain}
Since its introduction in 2008, blockchain technology has ushered in a new dimension of technology development, transforming industries from medicine and agriculture to finance and government. For instance, in 2016, IBM and the largest global ocean carrier, Maersk, launched a large-scale permission blockchain ecosystem called in TradeLens. Since then, TradeLens has gained overwhelming participation from the worlds’ top carriers, over 100 international ports and terminals, more than 10 government agencies, Standard Chartered, and numerous 3PL and 4PL companies \cite{ibm_news_maersk_2018,faridi_standard_2020,xia_keeping_2021}. DexFreight, a logistics startup in the trucking industry, has built a blockchain platform supported by smart contracts for food delivery. Using bitcoin, DexFreight’s blockchain allows the shipper and carrier to directly connect, negotiate rates, and schedule pickup and delivery \cite{obyrne_supply_2019}. In the food supply chain, IBM and Walmart have successfully launched a blockchain pilot program called Food Trust to check global food safety. This model has since been joined by companies like Nestle, Tyson Foods, Carrefour and Albertsons. 

Blockchains allow a set of mutually distrustful nodes to come to consensus on an immutable, append-only ledger. Key to this process is a consensus protocol. Some commonly used consensus protocols include Ripple, practical Byzantine Fault Tolerance (pBFT)\cite{castro_practical_1999}, Proof of Authority (PoA)\cite{noauthor_proof--authority_nodate}, Proof of Work (PoW)\cite{nakamoto_bitcoin_2008}, Proof of Stake (PoS), and Proof of Elapsed Time (PoET)\cite{chen_security_2017}, among many others.

One important aspect of blockchain is its support of smart contracts. Through smart contracts, businesses can automatically execute and verify contracts based on the fulfillment of certain predetermined conditions.
As a canonical example, blockchain based applications can ensure payment upon product delivery.
The ability for direct, secure, multiparty interaction in an untrusted setting is key to blockchain's widespread impact. Consequently, there is evidence that blockchain's ability to conduct automated negotiations according to each party's interests can significantly impact applications which utilize peer-to-peer trading or online auctions \cite{andoni_blockchain_2019,wang_secure_2020}. In this way, blockchain has the capacity to dramatically transform business research and operations across a variety of industries and settings.

\subsection{Some Challenges to Widespread Adoption of Blockchain Technology}
In business however, several barriers to the adoption of blockchain technologies remain. 
One of the main technical challenges of blockchain systems is \textit{interoperability}, the ability for disparate systems to communicate and transact with each other while maintaining their separate advantages. Verifying the existence of items in another ledger \textit{without} replicating the entire chain presents a fundamental challenge in this space. Further, the transfer of goods between two chains can neither create nor destroy assets on either chain - systems which achieve this are equivalent to a ``2-in-1" blockchain containing both ledgers \cite{lafourcade_about_2020}. Interoperability is an active area of research, with a wide variety of approaches \cite{perryman_blockchain_2020,kim_blockchain_2019,dale_one_2020}. A more detailed survey of the interoperability problem is covered by Belchior et al. \cite{belchior_survey_2020}.

The difficulty and cost of developing secure smart contract systems is an additional barrier to industry adoption of blockchain technology \cite{sara_rouhani_security_2019, prewett_blockchain_2019}. Software design flaws in smart contracts can compromise the safety of the entire system. There have been recent advances in automated generation of test cases, models for formal verification of security properties, and methodologies for practical security analyses for smart contracts. Nonetheless, several high-profile failures to achieve safety have resulted in some negative perception of the safety of blockchain systems \cite{tsankov_securify_2018,huang_smart_2019,byung_kim_automated_2020}. These concerns for safety have been amplified by existing bottlenecks in the throughput of several blockchain systems \cite{fan_performance_2020, prewett_blockchain_2019}.

Additionally, some consensus protocols including PoS, PoET, PoW and PoA have the potential for a submitted and committed transaction to be reversed. This problem is present to varying degrees in many peer-to-peer public consensus protocols. It is undesirable in many practical use-cases, leading some to prefer protocols like pBFT. While pBFT faces somewhat worse communication complexities, it is still preferred over others for its permanence \cite{de_angelis_pbft_2018}. A more in-depth analysis of the barriers to the adoption of blockchain technology can be found in \cite{kouhizadeh_blockchain_2021} and \cite{biswas_analysis_2019}.

There have been efforts to standardize blockchain implementations, providing simple examples and easy to access guides on designing and implementing smart contracts. For instance, Ethereum and Hyperledger have quick start examples that enable blockchain emulation from scratch within minutes. Nevertheless, various business applications of blockchain technology require specialized design of secure smart contracts and in-depth verification to ensure stability.
This leads to a dilemma for businesses, where in order to justify development of a system, they may have to build it, and to build it, they may have to justify the development cost. While emulators are potentially useful for determining bottlenecks in specific blockchain systems, a different solution may be needed to support broader feasibility analysis. Talaria is a blockchain \textit{simulator} that addresses the difficulty of justifying the business cost of developing and adopting a particular blockchain application. 

\section{Blockchain Simulators}
While some existing blockchain research focuses on low-level details, other research focuses on developing implementations for a particular application. 
This work explores the development of a system called Talaria (\href{https://github.com/Jiali-Xing/Talaria}{Talaria}), which departs from both paradigms by enabling a wide range of feasibility analyses. Talaria serves as a framework for assessing blockchain implementations in real world applications. 
Moreover, Talaria uses \textit{pre-existing statistical distributions} and \textit{event based scientific computing} back-ends to simulate the creation, transfer, and, importantly, processing of arbitrary transactions, which gives enhanced efficiency and accuracy. Talaria is by no means the first attempt at developing a blockchain simulator; Talaria builds on pre-existing simulator systems and addresses several gaps, as described in \autoref{fig:simulator_comparison}.

SimBlock \cite{aoki_simblock_2019, banno_simulating_2019} is a popular simulation platform. Given network and blockchain parameters, it simulates public blockchain systems which use PoW consensus.
Since it does not actually perform the mining procedure, SimBlock can easily scale to 10,000 nodes or extend to other blockchain protocols. While SimBlock is designed to have nodes propagate blocks, it does not simulate the transactions which are a key component in the assessment of a blockchain application. For example, supply chain or banking simulators require different transaction loads and put different requirements on the blockchain nodes. Moreover, SimBlock does not have native support for permissioned blockchains, which is important in many industry use-cases.

The Visualizations of Interactive, Blockchain, Extended Simulations (VIBES) simulator is another noteworthy blockchain simulation platform with many useful qualities \cite{stoykov_vibes_2017}. VIBES was explicitly designed to ``derive empirical insights on the behavior of a specific [blockchain] system.'' To that end, it has been designed to simulate a network with thousands of nodes and, as in most simulators, uses ‘fast-forward computing’ to provide the results of transactions spanning hours or days at three to five times their actual speed. The designers of the VIBES blockchain offer Proof of Work and Proof of Stake examples. 

However, VIBES achieves scalability at the expense of modularity; the system requires a relatively complicated, consensus protocol dependent fast-forwarding mechanism. 
Thus, this system seems essentially analogous to the BlockSim simulator extended by Talaria and BlockSim-Net, with BlockSim using the more accessible built-in event simulator from SimPy \cite{matloff_introduction_2008} in place of VIBES' Orchestrator module and complicated per-consensus-protocol fast-forwarding mechanism. As the field of consensus protocols and blockchain systems is rapidly evolving, the ability to incorporate new protocols and systems in an easy and efficient manner is critical to a simulator's ability to provide useful business insights.

The reader is referred to \cite{smetanin_blockchain_2020} for a complete survey of current blockchain evaluation methods. Motivated by these examples, an ideal blockchain simulator should have the following desirable properties. First, it should easily incorporate both public and private blockchain models. More generally, it should have built-in modularity, and layers of abstraction between the consensus, network, and application layers. Second, to ensure scalability it should not actually perform costly computations, such as mining for Proof of Work (PoW), but instead rely on statistical inputs for network latencies and transaction processing delays, block validation, and consensus protocol delays. Third, it should be easily extensible and user friendly. 
This is by no means an exhaustive list of desirable properties, but highlights the important qualities that distinguishes Talaria.

In the following sections we examine the technical details of Talaria and demonstrate its usefulness in feasibility analysis. Specifically, Talaria's applications are demonstrated within the context of a supply chain for public sector resource distribution scenario. In this domain, Talaria can provide performance metrics including estimates of reductions in cost. For businesses looking to implement a blockchain solution for their own supply chains, Talaria offers a solution to experiment and configure an efficient blockchain-supply chain strategy for a given operational environment.



\section{Talaria, the Permissioned Blockchain Simulator}
\begin{table*}
    \centering
    \caption{State of the art blockchain simulators - note that several simulators that are functional replicas of the simulators tabulated below are omitted. *Adversaries modeled $\dagger$ Simplified PoA implementation $\ddagger$ Transaction-level Simulation$\dagger\dagger$ Ability to interface with external simulators}
    \label{fig:simulator_comparison}
\begin{tabular}{c|c|c|c|c|c|c|c|c}
\hline
     \rowcolor{Blue}\textbf{
     Platform}&\textbf{Language}&\textbf{ Blockchain(s)}&\textbf{Consensus}&\textbf{ Adv* }&\textbf{Tx} &\textbf{Parallel-} &\textbf{Modular}&\textbf{Inter-}\\
     \rowcolor{Blue}&&&\textbf{Protocol}&&\textbf{Sim$\ddagger$}&\textbf{izable}&&\textbf{face$\dagger\dagger$}\\
     \rowcolor{LightPink}
     SimBlock\cite{aoki_simblock_2019}& Java &Public & PoW &N &N&N &N&N
     \\ 
     \rowcolor{LightCyan}
     BlockSim \cite{faria_blocksim_2019}& Python &Public&PoW,PoA&N&Y&N&Y&Y\\
      \rowcolor{LightPink}
     VIBES\cite{stoykov_vibes_2017} & Scala, React &Public&PoW&Y&Y&Y&N&N\\
     \rowcolor{LightCyan}
     Bitcoin Sim&C&Public&PoW&Y&N&N&N&N\\
     \rowcolor{LightPink}
     BlockSim-Net\cite{agrawal_blocksim-net_2020}&Python&Public&PoW&N&N&Y&Y&Y\\
     \rowcolor{LightCyan}
     Foytik et al\cite{foytik_blockchain_2020} & Unknown & Public & Raft& N & N&Y&Y&N\\
     \rowcolor{LightPink}
     Shargri-La\cite{okanami_shargri-_2020}&Rust&Public&N/A&N&Y&N&N&N\\
     \rowcolor{LightCyan}
     LUNES\cite{rosa_agent-based_2019}&Unknown&Public&PoW&Y&N&Y&N&N\\
     \rowcolor{LightPink}
     Talaria & Python &Public, &PoW, PoA$\dagger$ ,&Y&Y&N&Y&Y\\
     \rowcolor{LightPink}&&Private&PoS, PoET, pBFT&&&&&\\
     \hline
\end{tabular}
\end{table*}


Talaria is built in python on SciPy's event simulation framework, providing the capacity to simulate public and private blockchain systems in an efficient manner.

\subsection{Extensions to BlockSim}
    \label{section:expansions}
    Although it originally only captured the PoW models of Bitcoin and Ethereum, BlockSim is designed with a general blockchain framework in mind. Talaria extends the BlockSim work initiated in \cite{faria_blocksim_2019}, and to the authors' best knowledge, provides the first pythonic, modular permissioned blockchain simulator that can easily incorporate novel consensus protocols. 
    
    For demonstration purposes, we preselected two appropriate, popular, and powerful protocols for our implementation of permissioned blockchain: Clique \cite{noauthor_clique_nodate} (a specific instance of PoA) and pBFT. Two dominant Byzantine fault tolerant blockchain consensus protocols are PoA and pBFT \cite{de_angelis_pbft_2018}, with different strengths and trade-offs: PoA algorithms favor availability over consistency (as opposed to pBFT), and has better asymptotic communication complexity. On the other hand, pBFT has theoretical guarantees of strong consistency and finality \cite{de_angelis_pbft_2018}.
    
 \subsection{Periodical Transactions Broadcasting}
  \label{subsection:tx}

    Previously, BlockSim relied upon a transaction factory which generated transactions from randomly chosen origins/nodes, and then broadcasted them in bunches to all blockchain nodes. This design is, nevertheless, not able to take advantage of certain pattern of input requests. Therefore Talaria extends the transaction factory to read a \lstinline{JSON} file as input, where the {(node, number of transactions)} tuple is provided, as in \autoref{fig-JSON}.
    
    \begin{table}[h]
\centering
\caption{Input Information of Transactions}
\label{tab:tx-input}
\begin{tabular}{l|l|l|l|l}
\hline
\rowcolor{Blue}
\#Transactions & Day 1 & Day 2 & $\cdots$ & Day 180 \\ 
\rowcolor{LightPink}
Node 1         & 5     & 8     & $\cdots$ & 10      \\
\rowcolor{LightCyan}
Node 2         & 14    & 33    & $\cdots$  & 31      \\ 
\rowcolor{LightPink}
  $\cdots$ & $\cdots$  & $\cdots$  & $\cdots$ & $\cdots$  \\ 
\rowcolor{LightCyan}
Node 111       & 112   & 78    &  $\cdots$  & 97      \\ \hline
\end{tabular}
\end{table}
    
    According to the paper and comments in the code \cite{faria_blocksim_2019}, the original BlockSim authors planned to implement periodic broadcast of transactions. However, there was a minor issue in the implementation of transaction propagation in which transactions could be broadcast before any block was generated. This issue potentially prolonged the transaction list, drained computer memory, and was also not consistent with enterprise logic. Hence, Talaria has fixed the transaction propagation process, and makes it periodic. Though the transactions are currently generated based on a fixed time interval, this can be easily extended to any given distribution.

  \subsection{Discontinuous Simulation}
\label{subsection:discontinuous}
    Prior to Talaria, BlockSim ran for a continuous period of time with specified duration, regardless of whether the given set of transactions has been completely processed or not. Therefore, the simulation either  ends without processing all transactions, or keeps generating empty blocks even after finishing the processing of all transactions. Hence, to make the system more intuitive and appropriate for supply chain settings, Talaria extends BlockSim to stop simulation when all transactions have been processed. 

    Furthermore, Talaria extends BlockSim to support discontinuous simulations, e.g., when simulating sporadic transactions spanning multiple days. In real settings, the requests are not only periodical as described in the previous section, but also generated in separate time periods, such as 7 days with 10000 transactions per day. In most cases, the permissioned blockchain can finish processing a day's transactions within a couple of (simulated) minutes. Thereafter, instead of simulating the meaningless hours of heartbeats left for that day, Talaria saves the chains for each node and jumps to next day's simulation. As an artifact of the way Talaria reads in transactions, it considers all the transactions for a day to be finished if there are many consecutive empty blocks, e.g., 10.
    This allows the system to efficiently simulate a complex application over a large period of time, e.g., a supply chain spanning 100 days, without wasting time simulating empty blocks where there are no transactions to be confirmed.
    
  \subsection{Malicious Node Simulation}
    \label{subsection:malicious}
    The pBFT protocol can progress effectively as long as the fraction of Byzantine faulty nodes is below 1/3  \cite{castro_practical_1999} (In the following, we use malicious and faulty interchangeably). Talaria's implementation of the pBFT captures this fundamental characteristic of the protocol. As a permissioned blockchain simulator, Talaria implements the complete pBFT protocol and is, consequently, Byzantine fault tolerant. Nonetheless, in order to study the impact of Byzantine faults on the performance of the blockchain network, Talaria reads the Byzantine node type in the configuration file. Currently, any node can be assigned as either an active or passive Byzantine node as shown in the last column in \autoref{tab:node-input}. In the former case, the node will actively tamper with the consensus messages, resulting in invalid digests, while in the latter case the nodes drop arbitrary messages with a prescribed probability. Using this framework, the blockchain network resorts to view changes (re-election of the leader) if any leader is found to be actively or passively malicious.
    
    \autoref{tab:node-input} provides an example input to Talaria. Here, the second column is binary, denoting whether or not a node is an authority. Besides the binary value, the authority could also be implicitly determined by, for example, the location ID, such that every node with location ID smaller than 4 are authorities. In the last column of \textit{Byzantine Type}, $0$ indicates non-faulty nodes, while $1$ and $2$ are active and passive Byzantine respectively. 

    \begin{table}[htbp]
    \centering
    \caption{Example Input of Nodes}
    \label{tab:node-input}
    \begin{tabular}{l|l|l|l|l}
    \hline
    \rowcolor{Blue}
NodeID & Authority & Location     & Data & Byzantine \\ 
     \rowcolor{LightPink}
1      & 1         & Portland     &            & 2              \\ 
\rowcolor{LightCyan}
2      & 1         & Minneapolis  &            & 1              \\ 
     \rowcolor{LightPink}
3      & 1         & Honolulu     &            & 0              \\ 
\rowcolor{LightCyan}
4      & 1         & Yokohama     &            & 0              \\ 
     \rowcolor{LightPink}
5      & 1         & Hanoi        &            & 0              \\ 
\rowcolor{LightCyan}
6      & 1         & San Diego    &            & 0              \\ 
     \rowcolor{LightPink}
7      & 1         & Philadelphia &            & 0              \\
\rowcolor{LightCyan}
8      & 0         & Chicago      &            & 0              \\ 
     \rowcolor{LightPink}
9      & 0         & Pittsburgh   &            & 0              \\ 
\rowcolor{LightCyan}
10     & 0         & Newark       &            & 0              \\ 
     \rowcolor{LightPink}
11     & 0         & Vienna       &            & 0              \\ 
\rowcolor{LightCyan}
12     & 0         & Taipei       &            & 0              \\ \hline
    \end{tabular}
    \end{table}
    
    \subsection{Scalability}
    \label{scalability}
    As with all simulators, the scalability of the system is a concern. As with many simulators, Talaria runtime depends on the number of distinct events being simulated, including the number of messages sent and received, rather than the amount of simulated time. Future work can extend Talaria to process events in parallel, following some other works shown in \autoref{fig:simulator_comparison}.
    Thus, in some circumstances it is to be expected that the simulated time is less than or equal to the simulation run time. To see this in practice, see the pBFT subsection of the Case Study: Protocols section below.
    
    \subsection{Output Format}
    Inherited from BlockSim, Talaria outputs, in JSON format, a detailed report of the simulated propagation delays of transactions and blocks as well as the number and hashes of blocks in each node's chain. This report can be used to study various aspects of a permissioned blockchain system, such as the realized propagation delays between any pair of nodes or the performance and consistency of the blockchain.
    
\section{Case Study: Protocols}
  \subsection{Customized Consensus Protocols}
    \label{subsection:customized_consensus}
    Similar to the permissionless blockchains case, there are numerous consensus protocols that support permissioned blockchains, e.g., Proof-of-Stake \cite{king_ppcoin_2012}, Proof-of-Value \cite{noauthor_hybrid_nodate}, PoET \cite{chen_security_2017}, Proof-of-Reputation to name a few. As outlined in this work, we present two basic protocols: pBFT and a simplified version of PoA, since most of the aforementioned protocols can be derived using these two protocols. For instance, in the simplified version of Clique, the leaders are assumed to take turns to broadcast new blocks. It is straightforward to modify this order for PoET, where each node has a random waiting time and, consequently, the network has a random order of leaders based on elapsed time \cite{chen_security_2017}. Similarly, its pBFT implementation can be used to serve as a building block for simulating other complex protocols. For example, Tendermint \cite{buchman_latest_2019}, another BFT protocol, is inspired by and resembles the pBFT consensus procedure in that it has three communication steps in each round. Therefore, it would be natural to adapt the pBFT implementation to support a new Tendermint blockchain simulation.
    
  
\subsection{Proof of Authority (PoA)}
    \label{subsection:poa}
    First, Talaria implements a simplified permissioned blockchain simulator in the BlockSim framework with PoA consensus. Originally proposed by the co-founder of Ethereum and Parity Technologies, Gavin Wood, there is no clear boundary or universal definition for a PoA system \cite{noauthor_proof--authority_nodate}. One definition is to consider PoA as a special case of Proof-of-Stake, a consensus protocol that fairly distributes the right to validate blocks according to the size of stakes, where in PoA the validators' authority is the stake itself. In the scope of cryptocurrency systems, stake is the amount of the cryptocurrency actually owned by a node or group of nodes, but this notion can be generalized outside of this context.
    
    Rather than having miners race to solve an expensive block mining computation, block generation is left to the discretion of authorities. This mechanism guards against faulty leaders by the trust of authorities, i.e., malicious deviations from the protocol would risk a node's \textit{authority} and prevent it from being a leader in the future, \cite{noauthor_clique_nodate, de_angelis_pbft_2018}.
    
    \begin{figure}
        \centering
        \includegraphics[width=\columnwidth]{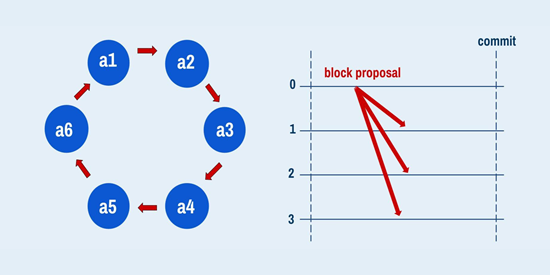}
        \caption{Our Simplified PoA Implementation}
        \label{fig:clique}
    \end{figure}
    
    
    Clique is a specific protocol in the PoA family, where a small, predefined group of leaders take turns in broadcasting new blocks, with only one round of message exchanges, i.e. block proposal, as opposed to the three stages of pBFT. Talaria's simplified PoA implementation reproduces the core concept of Clique with the assumption that there are no faulty nodes and thus no possibility of reversion of transactions \cite{noauthor_clique_nodate, de_angelis_pbft_2018}.
    
    Therefore, it reduces to the simplest scenario where every node in the authority group proposes blocks in turn, as shown in the left subgraph in \autoref{fig:clique}, and all other nodes accept blocks without further communication, as in the right subgraph in \autoref{fig:clique}. 
    
    Instead of a standalone blockchain production network, this simplistic model serves more as a starting point for further customization. Therefore, Talaria implements a PoA model with na\"ive network behavior, without faulty nodes.
    Given this framework, Talaria can extend to other popular and more complicated consensus algorithms used in permissioned blockchain settings such as the full-fledged Clique protocol. We now provide description of Proof-of-Elapsed-Time (PoET) and pBFT below.

  \subsection{Proof of Elapsed Time (PoET)}
  \label{subsection:poet}
    To demonstrate the extendibility of Talaria's simplified PoA system, a simple PoET protocol is implemented with 20 lines of additional code. Instead of mining, PoET blockchain system depends on random variables sampled from a exponential distribution in a hardware-protected environment to achieve a lottery-based block generating process  \cite{noauthor_hyperledgersawtooth-core_nodate, zhang_rem_2017}. In short, authorities wait for the time indicated by their random draws, and therefore, the entity with lowest random number gets to broadcast the new block. The full protocol thus differs from the simplified PoA only in terms of the leader selection process in each round, so it can be implemented with only minor modifications. A demonstration of the performance of this specific PoET implementation is not included, nevertheless, it illustrates the easy adaption of the aforementioned PoA abstraction, and justifies the extensibility of Talaria.
    
  \subsection{Practical Byzantine Fault Tolerance (pBFT)}
  \label{subsection:pbft}
  
  As mentioned above, the presence of a separate, equally valid and potentially conflicting version of the chain is called a \textit{fork}. If a consensus protocol does not allow for forks, the protocol is said to have \textit{finality}. For example, forks can occur in Nakamoto consensus when a miner mines a block containing a transaction, before learning about a longer chain including a different (potentially conflicting) block than the one to which it committed. The key to Nakamoto's 2008 analysis is that the probability of this happening to some particular block drops significantly, the longer a chain grows past that block \cite{nakamoto_bitcoin_2008}. 
    
    As finality seems a desirable property for many industry use-cases, Talaria implements a pBFT blockchain model following the seminal 1999 paper \cite{castro_practical_1999}. Here requests are a set of transactions, which contain several arbitrary strings, generated at specified nodes. After requests/transactions are collected, there are three consensus phases: pre-prepare, prepare, and commit. 
    
    
    \begin{figure}[h]
        \centering
        \includegraphics[width=\columnwidth]{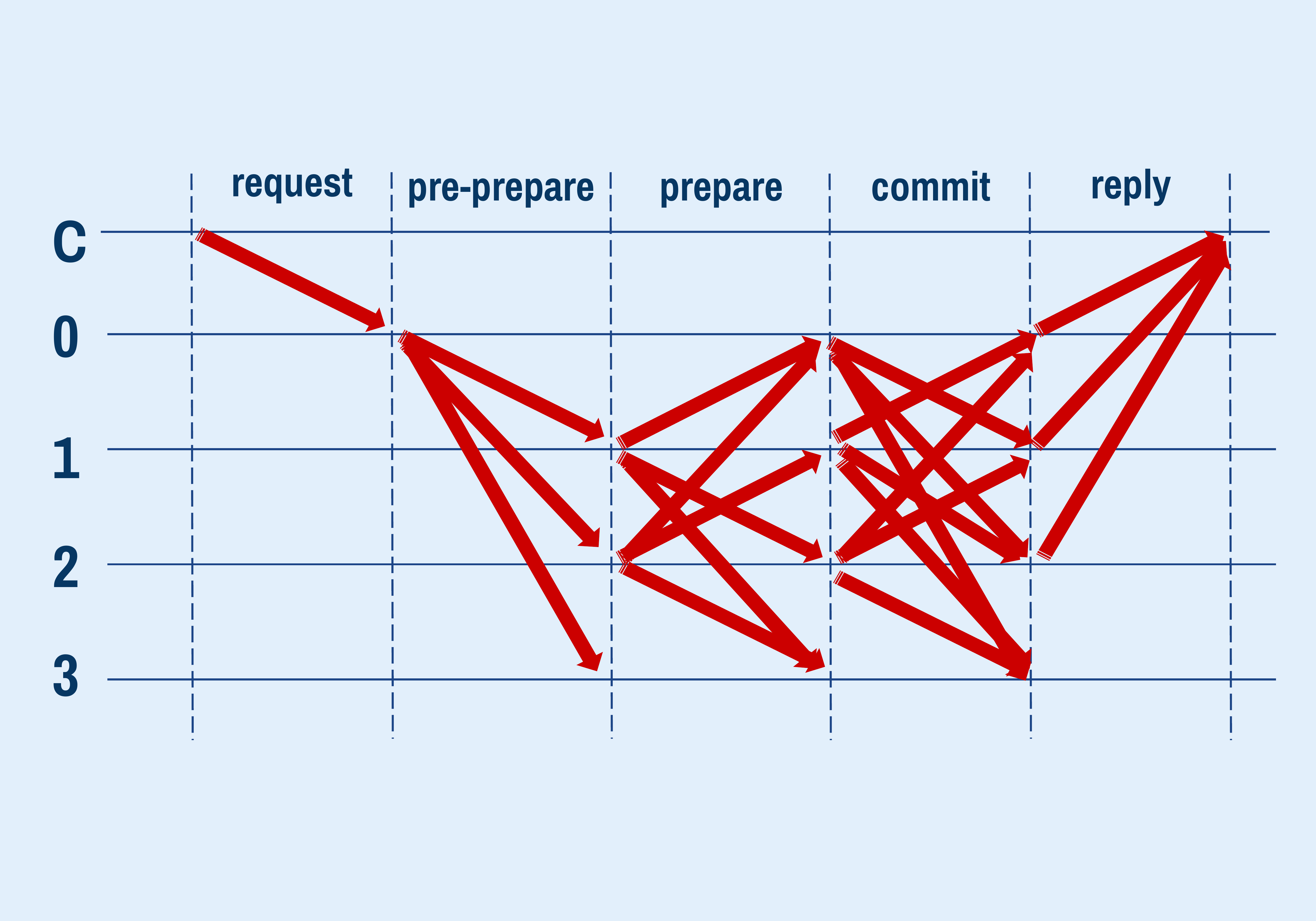}
        \caption{Normal Case Operation of pBFT with a faulty replica 3}
        \label{fig:pbft}
    \end{figure}
    
    In addition, BlockSim takes as input the statistical distributions of transaction and block validation delays, instead of actually running any type of cryptographical verification. In the pBFT consensus algorithm, Talaria specifies this delay with data from \cite{sukhwani_performance_2018}, where the author collected the statistics of \textit{time to process} incoming consensus messages using Hyperledger Fabric pBFT.
    
    In the presented demonstration, Talaria recovers the well known quadratic communication complexity of pBFT (or higher while recovering from a malicious leader). For the given implementation of pBFT, running the simulator with more than 21 authority nodes resulted in significantly higher running times. For instance, 22 authority nodes required an hour to finish, whereas 50 required more than eight hours. Note that increasing the number of follower nodes does not substantially increase the runtime of the simulator. This is because the number of messages required by each follower with respect to each block is linear in the total number of follower nodes and the number of authorities (i.e. followers can broadcast messages to every authority and each authority node will inform every follower of a finalized block). This can be improved to be sublinear with non-trivial changes. As Talaria is limited by the communication complexity of the consensus protocols under examination, this work presents the pBFT protocol as run on 13 authority nodes.
    
    
\section{Case Study: Supply Chains}
    
 \subsection{Use Case} 
Talaria may be employed in numerous applications. Among the many applications of blockchain, one promising use-case of much recent interest is in supply chain management. Quantifying supply chain costs (SCC) constitutes an area of research unto itself. All SCC metrics known to the authors of this paper include an entry for administrative costs. For instance, \cite{pettersson_measuring_2013} presents an example product and supply chain network for which administration alone is just under 20\% of full SCC, second only to manufacturing cost. Given this setting, a small but systematic reduction in the costs of these processes can lead to large aggregate savings for companies managing complex supply chains. Industrial interest in these expected cost savings is quickly emerging, as evident from pioneering projects such as \href{https://www.ibm.com/blockchain/solutions/food-trust}{IBM} and Maersk's \href{https://www.maersk.com/news/articles/2019/09/20/a-game-changer-for-global-trade}{TradeLens collaboration}, and Walmart’s FoodTrust \cite{noauthor_ibm_2021, madsen_game_nodate}. According to the United Nations Refugee Agency, the use of blockchain for voucher distrubtion in the Zataari Refugee Camp resulted in a 98\% reduction in administrative costs \cite{juskalian_inside_2018}. Moreover, the possibility of automated and dynamic management of a manufacturer's network of suppliers and customers presents the opportunity for savings in aspects of SCC outside of administrative costs.


%
\subsection{Defense Logistics Agency's Supply Chain}

To this end, in order to demonstrate Talaria, we focus on Defense Logistics Agency (DLA)'s supply chain. This is a very complex setting as defense supply chains may involve decisions for 10,000+ critical parts sourced from 1,000+ suppliers. 

In addressing these challenges Raytheon has developed the Defense Logistics Agency Supply Chain (DLASC) simulator under the DARPA Lagrange program. It models complex, dynamic, and high dimensional supply chains for the Department of Defense (DoD), and is used in conjunction with the Talaria permissioned blockchain simulator below. Specifically, Talaria can port output from the DLASC simulator; this is a good representative of SCM systems as DLASC is a very complex supply chain system.

 \subsection{Ecosystem: Connection Between Simulators}
    \label{subsection:SCM_simulator_connector}
\begin{figure}[htbp]
    \centerline{\includegraphics[width=\columnwidth]{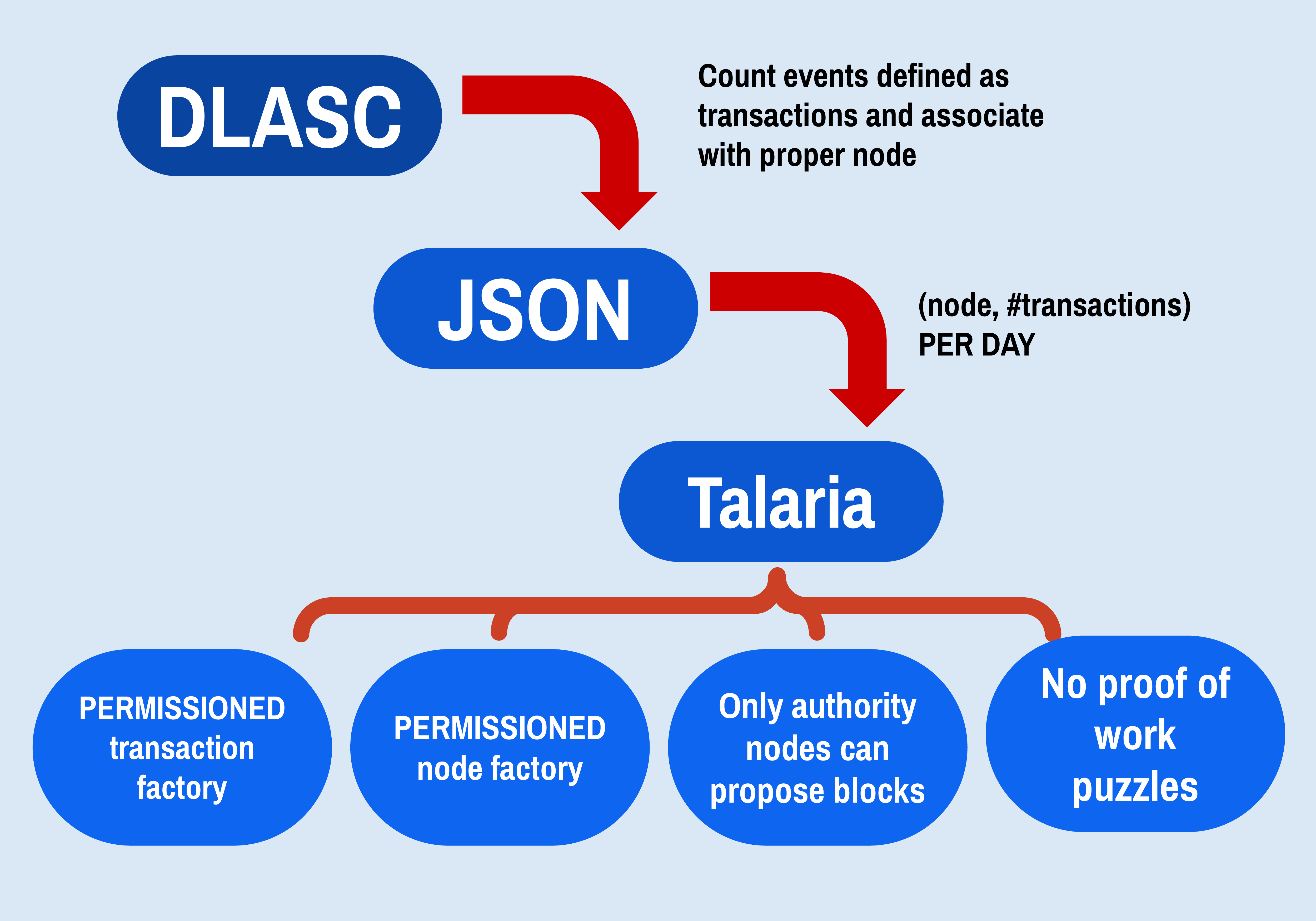}}
    \caption{Integration of Talaria with the DLA's Supply Chain Simulator}
    \label{fig-JSON}
    \end{figure}

    The Talaria simulator reads a \lstinline{JSON} file as input for transaction factory, which can come from another simulator. As an example, the presented work collects parts movement data from the Raytheon Technologies DLA supply chain simulator for modeling logistics performance, and pipes this information to be logged as transactions in Talaria, depicted in \autoref{fig-JSON}. 
 
    \begin{figure*}
        \centering
        \includegraphics[width=2\columnwidth]{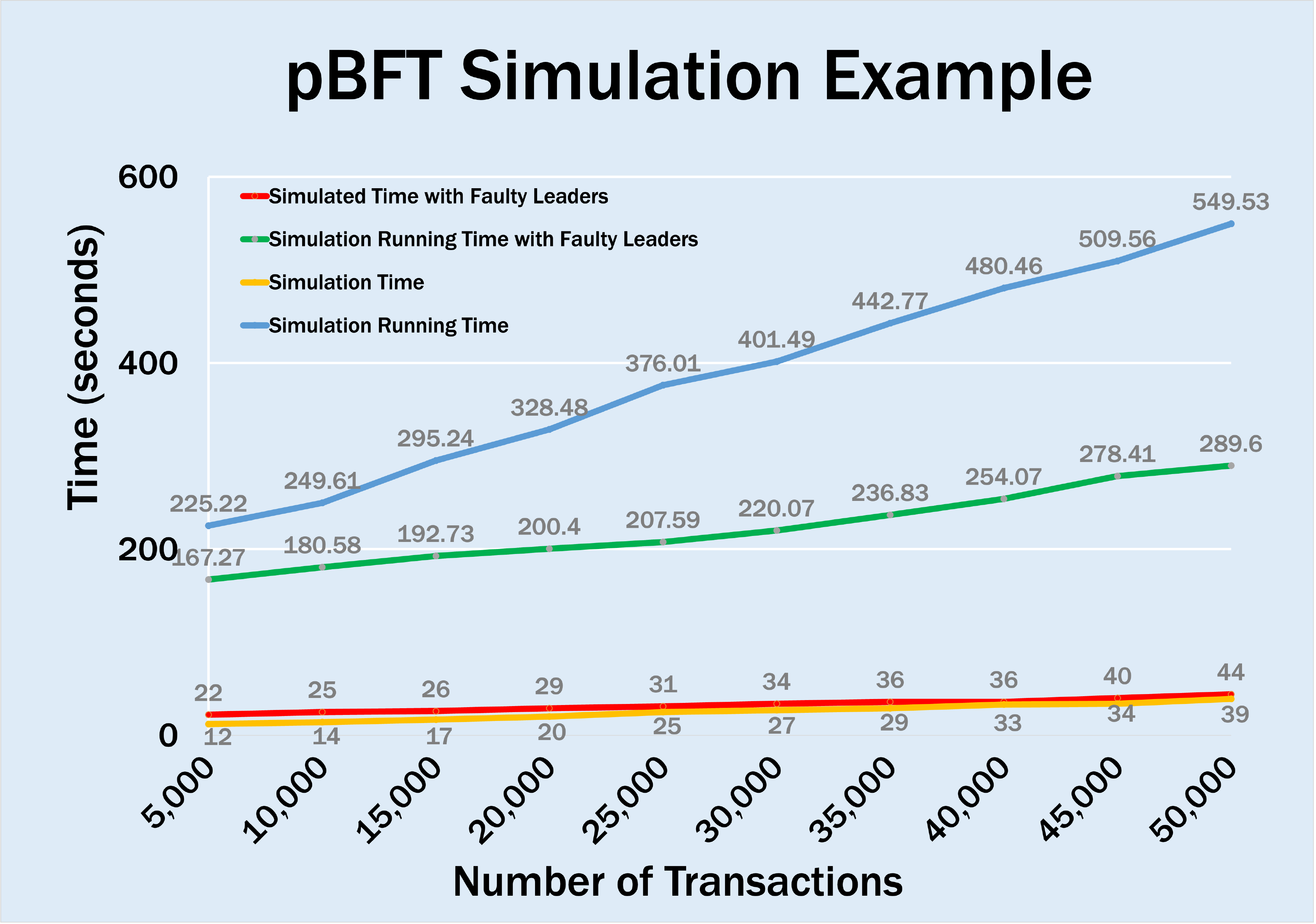}
        \caption{Example Simulation of pBFT run with 13 authority nodes}
        \label{fig:pbft_sim}
    \end{figure*}
    
The use case for blockchain simulators is to determine relatively quickly and with lower business and man-hour cost whether a blockchain solution is even feasible. This is distinct from that of blockchain emulators, which are more useful for analyzing specific metrics and aspects to some given blockchain implementation - such as smart contract security analysis.
    
\subsection{Results}
\begin{table*}[h]
\centering
\caption{Byzantine Fault Tolerant Result}
\label{tab:bft}
\resizebox{\textwidth}{!}{%
\begin{tabular}{l|l||ll|ll|ll|ll}
\hline
\rowcolor{Blue}
NodeID &
  Authority &
  \multicolumn{2}{l|}{Situation (1)} &
  \multicolumn{2}{l|}{Situation (2)} &
  \multicolumn{2}{l|}{Situation (3)} &
  \multicolumn{2}{l}{Situation (4)} \\
  \cline{3-10} \rowcolor{Blue}
 & 
  & 
  \begin{tabular}[c]{@{}l@{}} Byzantine \\
  Type\end{tabular} &
  \#Blocks &
  \begin{tabular}[c]{@{}l@{}}Byzantine \\ Type\end{tabular} &
  \#Blocks &
  \begin{tabular}[c]{@{}l@{}}Byzantine\\ Type\end{tabular} &
  \#Blocks &
  \multicolumn{1}{l}{\begin{tabular}[c]{@{}l@{}}Byzantine\\ Type\end{tabular}} &
  \#Blocks \\ 
     \rowcolor{LightPink}
1  & 1 & 1 & 1 & 2 & 3  & 2 & 15 & 1 & 13 \\ 
\rowcolor{LightCyan}
2  & 1 & 1 & 1 & 2 & 16 & 2 & 5  & 1 & 13 \\ 
     \rowcolor{LightPink}
3  & 1 & 1 & 1 & 2 & 8  & 2 & 5  & 1 & 13 \\ 
\rowcolor{LightCyan}
4  & 1 & 1 & 1 & 2 & 5  & 2 & 7  & 1 & 13 \\ 
     \rowcolor{LightPink}
5  & 1 & 1 & 1 & 2 & 7  & 0 & 29 & 0 & 13 \\ 
\rowcolor{LightCyan}
6  & 1 & 0 & 0 & 0 & 27 & 0 & 29 & 0 & 13 \\ 
     \rowcolor{LightPink}
7  & 1 & 0 & 0 & 0 & 27 & 0 & 29 & 0 & 13 \\ 
\rowcolor{LightCyan}
8  & 1 & 0 & 0 & 0 & 27 & 0 & 29 & 0 & 13 \\ 
     \rowcolor{LightPink}
9  & 1 & 0 & 0 & 0 & 27 & 0 & 29 & 0 & 13 \\ 
\rowcolor{LightCyan}
10 & 1 & 0 & 0 & 0 & 27 & 0 & 29 & 0 & 13 \\ 
     \rowcolor{LightPink}
11 & 1 & 0 & 0 & 0 & 27 & 0 & 29 & 0 & 13 \\ 
\rowcolor{LightCyan}
12 & 1 & 0 & 0 & 0 & 27 & 0 & 29 & 0 & 13 \\ 
     \rowcolor{LightPink}
13 & 1 & 0 & 0 & 0 & 27 & 0 & 29 & 0 & 13 \\ 
\rowcolor{LightCyan}
14 & 0 & 0 & 0 & 0 & 19 & 0 & 29 & 0 & 13 \\ 
     \rowcolor{LightPink}
15 & 0 & 0 & 0 & 0 & 19 & 0 & 29 & 0 & 13 \\ \hline
\end{tabular}%
}
\end{table*}

    \autoref{fig:pbft_sim} illustrates an example use case. 13 authority nodes with 3 faulty members are simulated. The two lines above are when the leader is benign, and thus does not trigger view changes, while the two lines below are when the first three leaders are all faulty, resulting in three successive view changes, which accounts for why the red dotted line is above the orange one.
    
    Another example of the output is shown in \autoref{tab:bft}, where 8868 transactions with 13 authority nodes of different Byzantine types are simulated. Consider the blockchain of the non-faulty nodes, i.e., those with Byzantine Type $0$, in situation (1) there are 5 malicious nodes actively tampering blocks, which violates the pBFT requirement of maximum of $f = 4$ faulty nodes. As a consequence, there is not enough consensus messages with valid digest, and thus no blocks are appended to legitimate nodes\footnote{However, faulty nodes can have ``enough'' prepare and commit messages because they incorrectly count themselves as benign.}. In situation (2), 
    5 passive Byzantine authorities drop messages with probability $40\%$. 
    Therefore, with probability $60\%$, the faulty nodes behave normally and help with block validation. This randomized behavior along with the fact that only 5 out of 13 authorities are faulty, makes it possible to complete all transactions among authorities. However, the non-authority nodes (the last two rows in the \autoref{tab:bft}) are not able to maintain a consistent chain with the authorities, as they need more than $2f = 8$ benign authorities to guarantee consistency, and thus, are left behind after block 19. Situations (3) and (4) are similar to situations (2) and (1) respectively, except the number of Byzantine nodes.
    In situations (3) and (4), because the number of Byzantine nodes are within the tolerance level $f = 4$, there are consistent blockchains among all benign nodes. Nonetheless, the randomness among Type 2 authorities prolongs the transaction procession. Because they are not completely faulty, they are not impeached as fast as the Type 1 Byzantine authorities. This example demonstrates the value of Talaria: it presents, including various corner cases, the performance and consistency of a blockchain application under given system and network conditions.

\section{Summary}
\label{section:conclusion}
    In this paper, we present a brief survey of existing blockchain simulators and outline their features. We also demonstrate Talaria, a new permissioned blockchain simulator with many potential applications. In particular, its application to support feasibility analysis in the permissioned setting can benefit companies and researchers in estimating the cost and performance of blockchain systems prior to the investment of resources in implementing a full fledged blockchain solution. Talaria supports simplified PoA, PoET, and pBFT consensus protocols out-of-the-box. Moreover, Talaria provides desirable features such as periodic transactions broadcasting, multiple-day simulation, passive or active malicious authorities, and transaction simulation. In future work Talaria can be extended to implement a parallelized blockchain simulator, as in BlockSim-Net.

\section{ACKNOWLEDGMENT}
This work was supported by the Defense Advanced Research Projects Agency (DARPA) and Space and Naval Warfare Systems Center, Pacific (SSC Pacific) under contract number N6600118C4031.

\printbibliography

\end{document}